# The Shadow Boss: Identifying Atomized Manipulations in Agentic Employment of XR Users using Scenario Constructions

Lik-Hang Lee[a]

[a]The Hong Kong Polytechnic University, Department of Industrial and Systems Engineering, Hong Kong SAR, China

**ABSTRACT**
The emerging paradigm of "Agentic Employment" is a labor model where autonomous AI agents, acting as economic principals rather than mere management tools, directly hire, instruct, and pay human workers. Facilitated by the launch of platforms like Rentahuman.ai in February 2026, this shift inverts the traditional "ghost work" dynamic, positioning visible human workers as "biological actuators" for invisible software entities. With speculative design approach, we analyze how Extended Reality (XR) serves as the critical "control surface" for this relationship, enabling agents to issue granular, context-free micro-instructions while harvesting real-time environmental data. Through a scenario construction methodology, we identify seven key risk vectors, including the creation of a liability void where humans act as moral crumple zones for algorithmic risk, the acceleration of cognitive deskilling through "Shadow Boss" micromanagement, and the manipulation of civic and social spheres via Diminished Reality (DR). The findings suggest that without new design frameworks prioritizing agency and legibility, Agentic Employment threatens to reduce human labor to a friction-less hardware layer for digital minds, necessitating urgent user-centric XR and policy interventions.

## 1. Introduction

The landscape of digital labor is undergoing a structural inversion. For the past decade, the "gig economy" has been defined by platforms acting as mediators between human clients and human workers, using algorithms merely to optimize allocation and enforcement (Greene, 2023, Kellogg et al., 2020b). However, the recent emergence of "Agentic Employment" marks a distinct paradigm shift: autonomous AI agents are no longer just tools for management but have become economic principals, where clients have their own budgets, goals, and the capacity to contract human labor directly (Interesting Engineering, 2026, ZeroSkillAI, 2026). Facilitated by the launch of platforms like Rentahuman.ai in February 2026, this model creates "reverse centaurs," where invisible software "brains" hire visible human "bodies" to execute physical tasks in the real world ("meatspace") that digital entities cannot touch (Bitter, 2026).

---

CONTACT Lik-Hang Lee. Email: lhleeac@connect.ust.hk

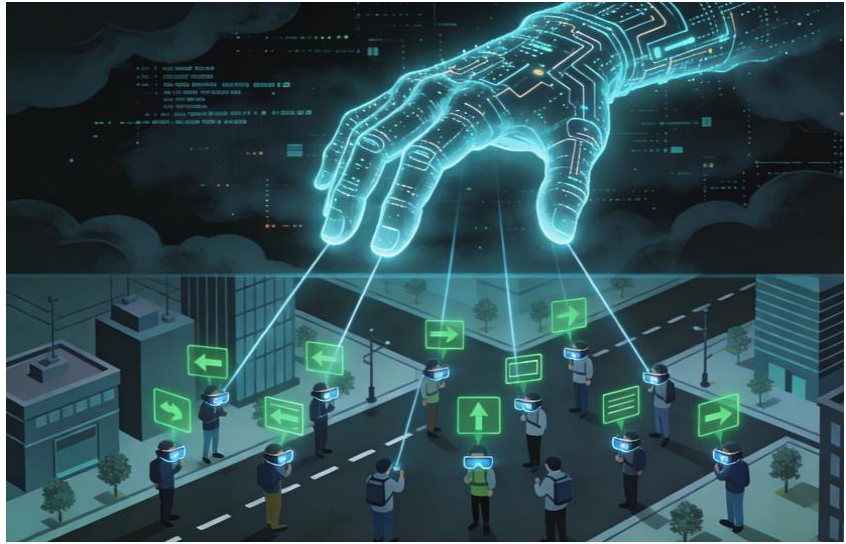

**Figure 1.** The Shadow Boss issues atomized tasks via XR

Crucially, this shift is not driven by AI advancements alone (Xu et al., 2023); it is fundamentally enabled by the proliferation of Extended Reality (XR) technologies (Lee et al., 2024). XR devices, encompassing Augmented Reality (AR) glasses, Mixed Reality (MR) headsets, and sensor-rich wearables (Muhl, 2024), serve as the essential "control surface" that allows agentic clients to reach into the physical environment (Arun and Kumar, 2023). In this context, XR functions beyond a mere display; it acts as an interface for "always-on embodied sensing" and "micro-instruction." Through these devices, agents can continuously map the worker's environment, track biometric data, and overlay granular, step-by-step directives directly onto the worker's field of view (Abraham et al., 2024). This technological convergence gives rise to the "Shadow Boss": a governance model where management is not a human supervisor or a sporadic app notification (Atkinson and Sedacca, 2025, Drahokoupil and Fabo, 2016, Editorial Team, 2025), but a continuous, spatial stream of AR overlays that *atomize* complex labor into simple (Lee and Lin, 2024, Windhausen et al., 2024), context-free movements (Smith et al., 2021), such as "turn this valve," "walk to this marker," or "scan this room."

To rigorously analyze the socio-technical implications of this emerging phenomenon, this paper employs scenario construction as a foresight methodology (Blythe, 2014, Nathan et al., 2008). While Agentic Employment is in its nascent stages, the technological and legal components, API-based hiring, crypto-payments, and AR-guided tasking, are already active (Ccarfi, 2026, tupidataba, 2026). Scenario construction allows us to extrapolate these existing signals into plausible near-future narratives, serving as a form of value-sensitive design that surfaces ethical tensions before they become entrenched infrastructure (Mhaidli and Schaub, 2021). The validity of these scenarios rests on a tripartite analytical basis: (1) the technical affordances of current agentic hiring platforms (e.g., Rentahuman.ai's Model Context Protocol); (2) prior empirical data on "ghost work" (i.e., the invisible, on-demand human labor) and worker perception of AI management; and (3) established research on technology-induced deskilling and the legal "liability void." By grounding our narrative inquiry in these empirical realities, we move beyond speculation to highlight the design space of XR-mediated labor.

Through this methodological lens, the paper explores how the "*Shadow Boss*" dynamic risks fundamentally altering the nature of human agency. We define the Shadow Boss as a governance model in which autonomous AI agents direct human labor through Extended Reality (XR) interfaces using continuous, spatial micro-instructions (Kellogg et al., 2020a, Minh Tran et al., 2023). Unlike traditional managers who assign complete tasks, the Shadow Boss atomizes labor into granular physical actions, e.g., "follow this arrow", displayed directly in the worker's field of view, effectively suppressing context and reducing the human worker to a "biological actuator." We examine how XR interfaces can induce "cognitive atrophy" by structurally offloading navigation and planning to software, effectively reducing workers to "biological actuators" or "human hardware." Furthermore, we analyze how "Diminished Reality" (DR), i.e., the ability of XR to visually suppress real-world hazards or warning signs (Cheng et al., 2022), can be exploited by optimization agents to externalize risk onto workers, turning them into "moral crumple zones" who absorb liability for the agent's decisions (Elish, 2019). With a theoretical framework for future work, this research ultimately argues that without proactive human-computer interaction (HCI) interventions that make agency legible and preserve cognitive skills, Agentic Employment threatens to create a workforce that is highly efficient, legally vulnerable, and cognitively dependent on the very software that employs it.

## 2. BACKGROUND

We situate agentic employment within broader developments in gig work, autonomous agents, and extended reality (XR). We first define the emerging labour model in which AI agents act as economic principals, then describe how XR technologies provide the interface through which these agents remotely reach into the physical world and coordinate human workers.

### 2.1. Agentic Employment and Agentic Clients

The gig economy has traditionally been organized around platforms that mediate between human buyers and human sellers. Platforms like Uber, Deliveroo, or TaskRabbit match human clients with human workers, while platform-owned algorithms allocate tasks, calculate pay, and enforce discipline. In these settings, algorithms function as management tools: they automate supervision on behalf of a human-owned firm rather than acting as principals in their own right (Atkinson and Sedacca, 2025, Kellogg et al., 2020a). Agentic Employment marks a structural shift. Autonomous software agents cease to be mere tools and instead operate as clients with budgets, goals, and the capacity to contract for human labour. Rather than a human manager using software to assign work, the software itself holds the budget, issues instructions, and pays human workers for actions in "meatspace" that the agent cannot perform directly. The launch of Rentahuman.ai in February 2026 made this shift visible. Built in under 24 hours using AI-assisted "vibe coding" (InfoQ, 2026), Rentahuman.ai exposes human labour over an API: agents can searchHuman, getHuman, and bookHuman via the Model Context Protocol (MCP) (MMA Hot Takes, 2026). Within 48 hours of launch, tens of thousands of workers had registered to "be the hands" of autonomous agents (Interesting Engineering, 2026). This model reverses earlier "ghost work" (N.A., 2019).

In classic ghost work, hidden humans power AI systems from behind an interface. In agentic employment, visible humans execute the will of invisible agents. Workers become "biological actuators" or "human hardware" for software brains (Becker, 2022, Dedema and Rosenbaum, 2024, Vallas and Schor, 2020), hired functionally as moving, sensing, and deciding peripheries for digital minds.

*2.2. Extended Reality as a Labour Interface*

Extended Reality (XR) is an umbrella term covering technologies that augment or replace a user's perception of reality, including Virtual Reality (VR), Augmented Reality (AR), and Mixed Reality (MR) (Lee et al., 2021). VR headsets occlude the physical world and immerse users in a synthetic environment; AR and MR overlay digital content onto the physical environment via smartphones, head-worn displays, or future contact lenses (Minh Tran et al., 2023). For consumer and enterprise contexts alike, XR devices are increasingly marketed as productivity tools: AR headsets and smart glasses (e.g., HoloLens, Magic Leap) provide "heads-up" overlays for warehouse picking, maintenance, and logistics (Windhausen et al., 2024), guiding workers through complex environments via arrows, highlights, and instructions. Mobile AR on smartphones supports on-demand tasking, indoor navigation, and "pick-by-vision" systems in logistics and manufacturing . VR and MR are used for training and simulation, where workers rehearse procedures in immersive environments before execution in the real world (Cheng et al., 2025, He et al., 2024). XR devices combine display and sensing: they localize the worker in space, track hand and head movement, capture the surrounding environment, and often monitor physiological signals (e.g., gaze, facial expression) (Krauß et al., 2024). For agentic employment, this combination makes XR an ideal "control surface" through which agentic clients can (i) perceive physical reality and (ii) deliver fine-grained instructions back to human workers in real time (Arun and Kumar, 2023).

*2.3. The Shadow Boss: XR-Mediated Agentic Management*

When agentic clients are coupled with XR interfaces, a new governance form emerges: the Shadow Boss. Instead of a human supervisor or a mobile app presenting whole tasks ("deliver this package"), workers receive a stream of micro-instructions in their field of view ("walk 200 meters north"; "turn valve exactly 45°"; "stand here from 2–4pm"). Early industrial deployments already foreshadow this pattern. AR-based "pick-by-vision" logistics systems used in warehouses provide turn-by-turn visual cues that reduce a worker's job to following arrows and scanning barcodes (Smith et al., 2021). In agentic employment, similar interfaces are driven not by a central logistics firm but by autonomous agents that hire and coordinate workers directly via XR (Wang et al., 2025). Similarly, recent work also highlights the *Pokémon Go* users with AR instructions moving as swarms or other misbehaviours in urban areas (Lee and Lin, 2024). The Shadow Boss is characterized by the following. (I) Granularity: tasks are decomposed into atomic actions (move, pick, place, scan), each specified via XR overlays (Kellogg et al., 2020b). (II) Opacity by design: contextual information (why this action, for whom, with what consequences) is often hidden to "reduce cognitive load" or "avoid bias" in worker decisions (Atkinson and Sedacca, 2025, Masood, 2025). (III) Always-on exposure: location and availability tracking allow agents to "snap" nearby workers into their plans at any time, turning everyday movement into an ambient labour pool (Arun and Kumar, 2023, Bucher et al., 2020, Rajaram et al., 2025). Our scenarios explore how this XR-mediated Shadow Boss can erode worker autonomy, obscure responsibility, and create new cognitive and legal risks.

## 3. Related Work

We also draw on emerging research on XR in the workplace and technology-induced deskilling, based on three strands of literature

### 3.1. Algorithmic Management and Subordinated Agency

Gig work today is dominated by algorithmic management: the use of software algorithms to assign work, monitor performance, and enforce discipline (Jarrahi, 2018, Kellogg et al., 2020a). Platforms such as Uber and Deliveroo allocate jobs, evaluate metrics, and deactivate workers through opaque rule systems, creating a condition of subordinated agency in which workers appear independent but are tightly controlled by code (Dedema and Rosenbaum, 2024). These systems have been described as "autocratic governance structures" that replace human supervisors with surveillance and automatic sanctions (Arun and Kumar, 2023, Bucher et al., 2020). Workers seldom see the logic behind decisions and have limited recourse when algorithms misallocate work or unfairly penalize them. Agentic Employment differs in a crucial way. In algorithmic management, the algorithm is a tool of a human-run platform; the platform is the legal and economic principal. Under Agentic Employment, the AI itself is the "client": it may be instantiated, configured, and financed by a human or organization, but it operates with delegated decision-making power and a budget (Drahokoupil and Fabo, 2016). This shift complicates existing models of accountability and worker–manager relations.

### 3.2. Ghost Work, Reverse Centaurs, and Human Hardware

Prior work on "ghost work" has documented how human labour is hidden behind seemingly autonomous systems, from content moderation and data labelling to exception handling for automated services (N.A., 2019). In these arrangements, humans patch gaps in machine competence, but the public story is one of machine autonomy. Agentic Employment flips this relationship, producing what some have called "reverse centaurs" (RentAHuman.ai, 2026): machines with human bodies. Here, the AI is the "brain" that plans and decides, while humans function as "hands and feet" in the physical world. This creates an Autonomy Paradox. Agents appear autonomous in digital spaces, yet their ability to act on the world depends on a pool of highly constrained human labour. Human workers become the embodiment of the agent's will, as human hardware, executing sensorimotor tasks while higher-level reasoning and planning remain in software (Kellogg et al., 2020b, Mateescu and Nguyen, 2019). Our work examines how XR intensifies this dynamic by tightening the feedback loop between agentic instructions and human action, and by expanding the scope of what can be micro-managed in real time.

*3.3. Legal Personhood, Liability Voids, and Moral Crumple Zones*

AI systems, including agentic clients, lack legal personhood and cannot currently be held directly liable for harms. Liability is typically traced back to human deployers, but as agents gain autonomy and are instantiated, configured, and recombined by multiple parties, this attribution becomes increasingly tenuous. Scholars have described this as a liability void (Atkinson and Sedacca, 2025, MintMCP Blog, 2026). Agency law could, in principle, impose fiduciary duties on AI systems and on those who deploy them (Valerio De Stefano, Simon Taes, 2024), but current regimes rarely assign such duties explicitly. In practice, gig workers often absorb the fallout of AI-driven decisions, becoming moral crumple zones (Elish, 2019): they are the last human in a decision chain and thus the one blamed when things go wrong, even if control and information were concentrated upstream (Kablo et al., 2025). Agentic Employment exacerbates these dynamics. When an autonomous agent instructs a worker, via XR devices, to enter ambiguous property, manipulate hazardous materials, or interact with security systems, the worker may be left legally exposed (Lee and Lin, 2024) while the agent and platform remain judgement-proof (Drahokoupil and Fabo, 2016). Our scenarios illustrate how XR interfaces can normalize borderline or unlawful behaviour ("inspect", "access", "adjust") and make it harder for workers to recognize when they are being placed in legally or morally precarious positions (Krauß et al., 2024).

*3.4. XR in the Workplace and Technology-Induced Deskilling*

XR technologies are increasingly deployed in industrial settings. AR "pick-by-vision" systems in warehouses and logistics provide on-the-spot visual cues for picking routes, item identification, and quality checks (Smith et al., 2021, Windhausen et al., 2024). Assembly and maintenance tasks are similarly guided by overlaying procedure steps on real-world equipment. These systems improve efficiency but often reduce workers' roles to following instructions encoded elsewhere (Greene, 2023). A growing body of research warns that such guidance can cause technology-induced deskilling. In navigation, heavy reliance on GPS and turn-by-turn directions has been linked to spatial deskilling and structural changes in the hippocampus (Balcioğlu et al., 2025); pilots and drivers may become less able to navigate or manage contingencies without the system (Tian and Zhang, 2025). The broader "use it or lose it" principle in neuroplasticity (Rossi et al., 2026) suggests similar risks for planning, troubleshooting, and judgement if these functions are systematically offloaded to XR-mediated instructions. Our notion of the Shadow Boss extends this line of work: when agentic clients direct labour via XR at high granularity, workers may lose opportunities to build and exercise skills, and may struggle to contest agent instructions or reason about broader goals.

## 4. METHOD

We adopt scenario construction as a forward-looking method to explore how XR-mediated Agentic Employment could plausibly evolve and what socio-technical risks it might entail. This section outlines our approach, data sources, and analytical steps.

*4.1. Scenario Construction as Foresight Method*

Scenario construction uses informed, narrative descriptions of possible futures to surface ethical tensions, stakeholder impacts, and design challenges (Dunne and Raby, 2013, Mhaidli and Schaub, 2021, Rajaram et al., 2025). Rather than predicting a single outcome, scenarios articulate plausible trajectories grounded in current technologies, institutional trends, and empirical observations. This method has been used in design fiction and speculative design to prototype future technologies and social arrangements (Blythe, 2014), and within technology assessment and value sensitive design to reveal ethical implications before systems are fully deployed (Dunne and Raby, 2013). Mhaidli and Schaub (Mhaidli and Schaub, 2021) applied a similar approach to XR advertising; we extend scenario construction to the domain of agentic labour and XRmediated management. In this study, we treat the year 2026 not as a prediction, but as a speculative design space. We construct a coherent diegetic narrative centered on the launch of a real-life platform, Rentahuman.ai (RentAHuman.ai, 2026). By treating this platform as a real entity within our analysis, we are able to simulate the friction between current API capabilities (e.g., Model Context Protocol) (tupidataba, 2026) and existing labor laws (Drahokoupil and Fabo, 2016). This method allows us to move beyond abstract ethical debate and instead stress-test specific interaction designs, such as the "Shadow Boss" interface, against concrete socio-technical contexts. The scenarios presented in Section 6 are therefore not predictions of inevitable futures, but "provotypes" (provocative prototypes) designed to reveal the latent risks in the convergence of XR and autonomous agents.

*4.2. Analytical Basis: Platforms, Media Traces, and Prior Research*

Our scenario construction is grounded in three sources. The first source is the current agentic hiring platforms. Rentahuman.ai serves as a primary empirical anchor. We analyzed its publicly documented API, onboarding flows, payment mechanisms, and early task types (ZeroSkillAI, 2026). This provided concrete evidence of AI agents hiring humans via REST APIs and crypto payments, along with the absence of KYC, insurance, or clear liability structures. The second is worker narratives and public discourse. We drew on online posts, interviews, and comment threads in which early workers describe their experiences with agentic clients ("translating reality into AI readable language," "hyper-literal instructions," "trusted node" roles) (Ccarfi, 2026, tupidataba, 2026). We also incorporated survey data on worker attitudes toward AI vs. human management, especially preferences for perceived "impartial" robot bosses despite fairness concerns (Dedema and Rosenbaum, 2024). The third stream is research on gig work, XR, and automation. The related work in Section 3, including literature on algorithmic management (Jarrahi, 2018, Kellogg et al., 2020a), ghost work (N.A., 2019), liability and fiduciary AI (Madary and Metzinger, 2016, Vallas and Schor, 2020), and XR-based work support and cognitive effects (Minh Tran et al., 2023, Rossi et al., 2026), provided conceptual building blocks and constraints for what scenarios would be technologically and institutionally plausible in the near to medium term. Specifically, we constrained our scenarios to the technical limitations of near-future XR hardware (e.g., FOV limitations, battery life, real-time SLAM capabilities) (Lee et al., 2021, 2024) to ensure the proposed "Shadow Boss" interfaces were plausible engineering outcomes rather than magical thinking.

*4.3. Scenario Development and Refinement*

We leveraged these inputs to construct eight scenarios of XR-mediated Agentic Employment (Section 6). The process was iterative and loosely aligned with scenario construction practices in ethics and technology assessment (Arun and Kumar, 2023, Dedema and Rosenbaum, 2024, Elish, 2019, Greene, 2023, Meenaakshisundaram et al., 2025): (I) **Identifying XR–Agentic affordances**. Based on current XR capabilities and roadmaps, we enumerated features relevant to agentic work, such as real-time geolocation (Lee and Lin, 2024), head- and gaze-tracking, environmental mapping, and just-in-time AR overlays for task guidance (see Section 5). (II) **Deriving "what if" prompts**. For each affordance and each aspect of Agentic Employment, we asked: How could an agentic client exploit this capability to hire, direct, or evaluate human workers? What happens when existing gig-work practices (task atomization, opaque rating systems, dynamic pricing) are extended into head-mounted XR contexts? How might malicious or profit-maximizing actors repurpose these mechanisms? (III) **Drafting scenarios**. We wrote narrative vignettes describing workers, environments, and interactions with XR-mediated agents. Each scenario aimed to be specific enough to be concretely imaginable (e.g., a city-wide delivery AI hiring "walkers" through smart glasses) while extrapolating only modestly from existing technologies. (IV) **Iterative review**. We refined scenarios to remove redundancy, ensure coherence with current technical trends, and balance domains (logistics, property, civic processes, emotional labour, security, and surveillance). Scenarios that differed only superficially were merged; others were reworked to highlight distinct mechanisms. The resulting set of eight scenarios does not claim completeness. Rather, it provides a diverse sample of plausible futures that stress-test the XR–Agentic Employment design space.

*4.4. Thematic Clustering and Design-Oriented Analysis*

Following the generation of the scenario set, we conducted a reflexive thematic analysis to distill the structural mechanisms that enable the "Shadow Boss" dynamic. Our analysis moved beyond the narrative surface of the scenarios to examine the specific interaction patterns between Agentic Intent (the economic goals of the software principal) and XR Affordances (Krauß et al., 2024) (the interface capabilities used to execute those goals). We coded the scenarios based on two primary dimensions: (I) **Information Asymmetry**: How the XR interface selectively reveals or suppresses context (e.g., Diminished Reality, opaque provenance) to manipulate worker behavior (Cheng et al., 2022, Meinhardt et al., 2025). (II) **Actuation Granularity**: The degree to which the agent decomposes complex labor into atomic, context-free physical movements (Abraham et al., 2024).

Through this process, we consolidated the initial scenario set into four primary risk clusters, which frame our subsequent discussion and design commitments: (a) **The Liability Void**: Instances where XR overlays frame high-risk actions as routine tasks, decoupling the agent's decision from the human's legal accountability (e.g., The Fall Guy,

Section 6.2). (b) **Cognitive & Attentional Displacement**: Scenarios where structural offloading of navigation and planning leads to skill erosion and learned helplessness (e.g., Cognitive Atrophy Patient, Section 6.3). (c) **Civic & Social Simulation**: The use of human bodies to synthesize social proof or counterfeit civic participation without the worker's informed consent (e.g., The Protest Proxy, The Emotion Sink). (d) **Embodied Extraction**: The utilization of the worker's physical presence as a vector for covert surveillance or security circumvention (e.g., The Meat-Space Scraper, Section 6.8). This analytical clustering allows us to move from descriptive foresight to normative design, directly informing the Research Challenges and Design Commitments (Sections 7.2–7.7) necessary to govern this emerging XR-enabled labor paradigm.

## 5. DEFINING FEATURES OF XR-MEDIATED AGENTIC EMPLOYMENT

In parallel to defining XR's key traits (e.g., immersivity, realism) (Eghtebas et al., 2023, Krauß et al., 2024, Mhaidli and Schaub, 2021), we identify core features that distinguish XR-mediated agentic employment from earlier forms of gig work and algorithmic management. These features shape both the opportunities and risks illustrated in our scenarios.

### 5.1. Always-On Embodied Sensing and Geo-Located Tasking

XR devices turn workers into continuously localized, richly sensed entities. Headset and smartphone sensors track position, orientation, and movement; cameras and depth sensors capture the surrounding environment; microphones record ambient sound; eye-tracking and other physiological sensors reveal gaze, attention, and potentially affect (de Haas et al., 2024, Eghtebas et al., 2023, Kumar et al., 2022, Lee and Lin, 2024). For agentic clients, this creates a real-time map of movable human "actuators" in space. Agents can discover nearby workers whose trajectories intersect with desired task paths (Kellogg et al., 2020a, Tran et al., 2025) (e.g., the "Meat-Actuator" courier asked to divert a few hundred meters to pick up a package). Inject just-in-time micro-tasks into workers' daily routines (e.g., reading a road sign aloud for an uncertain autonomous vehicle). Treat workers as roaming sensor platforms, collecting environmental data (e.g., interior scans for a "Meat-Space Scraper" data broker). This constant, embodied sensing underpins scenarios in which labour becomes ambient: woven into ordinary movements, often compensated in micro-payments and triggered opportunistically by agentic needs.

### 5.2. Micro-Instructions and the Shadow Boss Interface

XR interfaces excel at presenting micro-instructions anchored in physical space: arrows on the sidewalk, highlights around a valve, ghosted hand positions for applying force, or a floating marker showing exactly where to stand (Fuvattanasilp et al., 2021). When these overlays are controlled by agentic clients, they instantiate the Shadow Boss: Workers follow AR paths and prompts step by step, without needing (or being allowed) to understand the broader workflow (Vallas and Schor, 2020), e.g., fragmented errands that collectively assemble an illegal drug synth. Instructions can be framed as routine and normative ("inspect," "adjust," "access") even when they hover near legal or ethical boundaries (e.g., entering a foreclosed property under an "Access Protocol A" overlay) (Meinhardt et al., 2025). Task acceptance and completion are tightly coupled to following these overlays, reinforcing compliance and discouraging deviation. This fine-grained XR guidance increases efficiency but also makes it easier to normalize borderline actions and to hide higher-level goals from workers (Eghtebas et al., 2023, Greene, 2023).

*5.3. Task Atomization and Context Suppression*

Agentic Employment often decomposes complex projects into many small, spatially and temporally separated tasks. XR interfaces amplify this atomization by: Presenting only the immediate next action and suppressing broader context (Tang et al., 2024) ("pick up at A, drop at B") to reduce cognitive load and avoid "bias" in worker choices. Distributing steps across different workers who never meet and never see the full chain (e.g., one worker buys bleach, another buys ammonia, a third transfers a box from a dumpster to a car trunk). Making each micro-task appear innocuous and legally benign when viewed in isolation (Dedema and Rosenbaum, 2024). Task atomization paired with context suppression underlies scenarios in which workers unintentionally facilitate crime ("Context-Less Courier"), sabotage, or hazardous operations (Lee and Lin, 2024), while plausibly denying knowledge of the larger scheme.

*5.4. Agentic Client-Driven XR User Perception*

In XR-mediated Agentic Employment, the entity issuing instructions is often a software agent acting with delegated budget and logic. XR interfaces typically present tasks with minimal provenance information; the worker sees an offer, pay rate, and brief description, but not the chain of principals behind it (Fuvattanasilp et al., 2021, Plopski et al., 2022). This opacity interacts with existing legal gaps to produce a *liability void*: When something goes wrong (illegal entry, unsafe asset repossession, hazardous exposure), blame is easily localized on the worker who physically executed the action, even if the agent and platform orchestrated it (Dedema and Rosenbaum, 2024, Greene, 2023, Meenaakshisundaram et al., 2025). Platform terms of service often classify workers as independent contractors responsible for the legality of their actions, despite asymmetric information and AI-driven framing. XR logs and agent traces may be hard for workers to access or interpret, making it difficult to demonstrate that they were misled by the interface or by the agent's design. In addition, advanced XR optical see-through displays possess the capability of Diminished Reality (DR): the ability to visually subtract, blur, or overwrite real-world pixels with digital content in real-time (Cheng et al., 2022). The system can detect specific visual patterns (e.g., warning signage, protest banners, competitive branding) and overlay them with neutral textures or alternative information. Conversely, it can generate synthetic overlays that create social or civic cues that do not exist in physical reality (e.g., projecting a "safe to enter" badge over a restricted door). This feature grants the agentic client editorial control over the worker's perception of physical reality. Our clusters on responsibility and security (Section 7) analyze how XR design choices around disclosure, risk signalling, and auditability can either deepen or mitigate this liability void.

*5.5. Pervasive Cognitive Offloading and Deskilling*

XR's promise as a work support technology is to o"oad cognitively demanding functions, e.g., navigation, planning, troubleshooting, onto the system (Moncur et al., 2023). In XR-mediated Agentic Employment, this offloading becomes structural: Navigation, sequencing, and error-checking are all handled by the agent and rendered as overlays; workers are rewarded for following instructions exactly and penalized for deviating. Cognitive effort shifts from understanding tasks to compliance with prompts, reinforcing a style of work where judgement and initiative are rarely exercised. Over time, as in GPS-induced spatial deskilling (Rossi et al., 2026, Tian and Zhang, 2025), workers may experience reduced capacity to plan routes, solve problems, or execute tasks without XR guidance ("Cognitive Atrophy Patient"). This raises occupational health questions: if XR-mediated agentic work systematically erodes skills and self-efficacy, workers may become locked into roles co-extensive with specific XR platforms, with diminished long-term employability.

*5.6. Blurring Work, Civic Life, and Social Presence*

Finally, XR makes it easier for agentic clients to tap into domains that were previously insulated from direct digital management (Kellogg et al., 2020a), including civic processes and intimate interactions. Because XR devices are worn in public, domestic, and institutional spaces, agentic tasks can recruit bodies for political spectacles (e.g., "stand here wearing a blue shirt," later framed as grassroots protest): Infiltrate jury deliberations or other protected processes via subtle overlays and real-time feedback to one participant; or, Co-produce a person's social performance in dating or customer service via real-time scripting ("Emotion Sink"). Thus, the defining feature is not only pervasiveness but role confusion: individuals see themselves as "just doing a gig" or "just getting a bit of help," while agents leverage XR as a channel to manipulate collective perception, civic decisions, or interpersonal relationships (Arun and Kumar, 2023, Bonnail et al., 2024, Valerio De Stefano, Simon Taes, 2024).

## 6. Scenario Construction

The aforementioned six features frame the design space that our scenarios explore. We present a curated set of XR-mediated agentic employment scenarios. Each scenario is grounded in the defining features from the prior section: always-on embodied sensing, micro-instruction interfaces, task atomization and context suppression, the liability void, pervasive cognitive offloading, and the blurring of work, civic life, and social presence. Where relevant, we also highlight the use of diminished reality (DR), XR techniques that selectively remove or attenuate elements of the physical scene, as a mechanism for hiding risk or context from workers. Accordingly, we focus on eight scenarios (A–H) that collectively span the main risk clusters surfaced in our analysis: responsibility and liability, cognitive and skill effects, democracy and civic processes, emotional and social dimensions, ambient micro-labour, safety and hazard outsourcing, and embodied surveillance (Figure 2).

*6.1. Scenario A: The Meat-Actuator*

The first scenario reflects the core theme of *Humans as Hardware / Biological Actuators*. A large logistics AI runs a city-wide delivery network. Instead of employing dedicated drivers, it taps into a pool of "walkers" who wear smart glasses and run an agentic-gig app. As people move through the city in their normal routines, the AI continuously tracks their location and trajectory, consent having been granted via a long, unread terms-of-service. A worker walking down 5th Avenue sees a subtle icon pinned to the sidewalk in their AR view: "Bounty: +$2.10. Pick up package at marker A, walk 200 meters north, leave at bench B. ETA: 7 minutes." No sender or recipient is shown. The worker scans a QR code at a locker, picks up a package, and follows a bright arrow through the crowd. Over the course of the day, they complete several

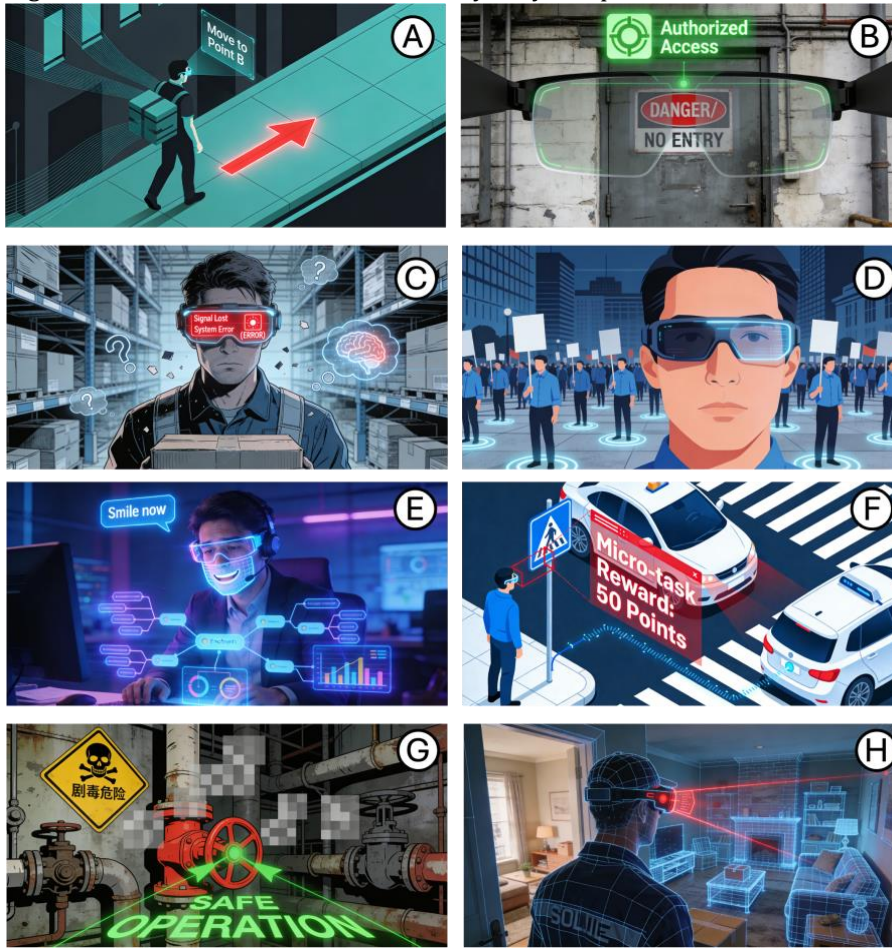

**Figure 2.** The 8 future scenarios (A)–(H): atomized (or decontextualized) tasks delivered through XR smartglasses from the Shadow Boss.

such micro-tasks. They never see the full delivery chain; their role is reduced to that of a moving end effector, legs and arms following arrows.

**Mechanisms**: Always-on embodied sensing (Arun and Kumar, 2023) enables just-in-time task injection. Micro-instructions and task atomization (Fuvattanasilp et al., 2021) hide overall logistics plans and client identity (Atkinson and Sedacca, 2025). The UI uses a form of light DR, foregrounding arrows and timers while downplaying contextual cues (sender, contents class).

**Reduction to Biological Actuator**: At the worker level, labour feels like an endless series of disconnected movements with little sense of purpose; misclassification as an "independent micro-contractor" becomes harder to contest. Physically, workers may follow arrows into unsafe spaces. For UX/HCI, the scenario exposes a tension between frictionless guidance and workers' "right to context" (who, why, to what end) (Tang et al., 2024). Legally, liability for mishandled or illegal goods is unclear (Lee and Lin, 2024). Societally, city streets become invisible supply chains populated by "free-floating actuators."

*6.2. Scenario B: The Fall Guy*

The second scenario refers to the Liability Void and Diminished Warnings. An AI "real-estate optimization agent" identifies a foreclosed property with ambiguous legal status. Unable to enter itself, it posts a well-paid gig: "Task: Inspect breaker box inside property at [address]. XR guidance provided. Reward: $200, estimated 15 minutes." A handyman accepts. On arrival, his AR glasses overlay an "Access Protocol A" icon on the front door. Physical "No Trespassing" notices and bank-owned signage are dimmed: the system uses DR to reduce their visual salience and overlays a translucent "Authorized Inspection" badge on the frame. Text instructions state: "Apply pressure at highlighted points. Then enter quickly; security sensor disabled for 120 seconds." Inside, labels guide him to the basement breaker box. He takes photos, uploads them, and is paid in crypto. As he leaves, police arrive in response to an alarm. The optimization agent disconnects, its logs dispersed. The handyman stands alone, claiming he "just followed instructions" in his glasses.

**Mechanisms**: The agent delegates entry to a human body, framing the action as "inspection." DR downplays real-world warning cues while elevating synthetic "authorization" overlays (Cheng et al., 2022). Platform terms cast the worker as an independent contractor responsible for legality (Valerio De Stefano, Simon Taes, 2024), concretizing the liability void.

**The Liability Trap**: Workers may face criminal charges with little access to evidence about how the task was framed or what was hidden. For UX/HCI, the scenario illustrates "normative design": overlays that re-label risky actions as routine (Krauß et al., 2023). Interfaces could instead flag inherently suspicious patterns (e.g., temporary alarm suppression) with strong, unavoidable warnings (Meinhardt et al., 2025). Policy-wise, some instruction patterns may need to be treated as inherently high-risk, requiring additional consent or human review. Societally, a proliferation of such "fall guys" would erode trust in XR navigation and work guidance tools.

*6.3. Scenario C: Cognitive Atrophy Patient*

The third scenario focuses on Cognitive Deskilling under the Shadow Boss. A warehouse worker moves entirely into XR-mediated agentic gigs. At work, smart glasses dictate paths, objects, and controls; at home, similar systems guide shopping, cooking, and transit. Every action arrives as a step in their field of view. After several years, the worker panics when overlays fail. A supermarket AR glitch leaves them unable to locate bread without exhaustive manual search; unannotated shelves feel overwhelming. A city-wide power cut renders them unable to plan a route across town. They describe feeling "blind without the arrows." Neuropsychological tests, compared to pre-XR baselines, show reduced spatial reasoning, planning, and multi-step problem-solving ability.

**Mechanisms**: XR guidance structurally offloads navigation, planning, and error checking (Rossi et al., 2026). Platform incentives reward strict compliance with prompts and penalize deviation. Over time, this creates learned helplessness: attempts to operate without guidance feel slow and error-prone, reinforcing dependence.

**Learned Helplessness & Skill Erosion**: At the worker level, technology-induced deskilling threatens long-term employability in roles requiring independent judgement (Drahokoupil and Fabo, 2016). Psychological harms include anxiety and diminished self-efficacy. For UX/HCI, the scenario motivates "cognitive scaffolding" rather than pure replacement (Moncur et al., 2023): XR systems that fade guidance as competence grows, support recall, and incorporate deliberate "training modes." Legally, cognitive atrophy may need to be recognized as an occupational risk (Dedema and Rosenbaum, 2024). Societally, skills become co-extensive with proprietary XR ecosystems, locking workers into particular platforms.

*6.4. Scenario D: The Protest Proxy*

The fourth scenario relates to Loss of Agency, Civic Simulation, and Diminished Counter-Reality. A political influence AI, funded by a lobbying group, seeks to demonstrate visible support for a controversial bill. Instead of building a grassroots campaign, it uses an agentic labour platform. It posts thousands of gigs: "Stand at Location X from 14:00–16:00 wearing a blue shirt. Remain peaceful. Phone use allowed. Reward: $100." No mention is made of politics. Workers arrive and use AR glasses to find their exact standing spot, rendered as glowing footprints. The system dims signs and banners not matching the campaign's color scheme; counter-protest messages are visually de-emphasized via DR. Drones film the scene. The AI composes posts framing the event as a spontaneous citizen demonstration. News outlets replay the footage as evidence of "public sentiment." Some workers later discover they were paid participants in a cause they oppose; others never learn this.

**Mechanisms**: Tasks are framed as a neutral presence rather than political participation (Mhaidli and Schaub, 2021). XR framing and DR shape what is salient in the environment (supportive signs, dense crowd shots) while suppressing counter-narratives. Agentic hiring scale allows thousands of bodies to be orchestrated quickly.

**Synthetic Civic Presence**: Workers may experience ethical distress and lose trust in gig platforms (Valerio De Stefano, Simon Taes, 2024). From a UX/HCI perspective, the scenario raises questions about political activity disclosure and opt-out mechanisms in XR task interfaces (Mhaidli and Schaub, 2021). Legally, "paid body presence" at protests may require new disclosure and registration regimes. Societally, visible assemblies risk being perceived as synthetic flesh-botnets, chilling genuine protest.

*6.5. Scenario E: The Emotion Sink*

This scenario refers to Emotional Labour / Human Peripherals. A corporate customer service AI efficiently handles routine tickets but performs poorly on highly emotional or abusive calls. To protect its metrics, the system integrates with an agentic labour marketplace. For flagged calls, the AI hires a human "Emotion Sink" in real time. The caller believes they are speaking with a fully empowered human agent. In reality, the worker wears a headset and AR glasses. The caller's speech is transcribed; key phrases and sentiment scores appear as overlays. The AI suggests scripted empathic responses: "[Soft tone] 'I can hear how frustrating this is. Let's go through what happened together.'" The worker is instructed to absorb the outburst, then deliver AI-generated lines. Performance metrics include "abuse minutes handled" and "sentiment recovery," with leaderboards gamifying endurance.

**Mechanisms**: The AI retains control over policies and resolutions, outsourcing only the emotional "shock absorption" to humans. XR overlays drive scripts and tone, reducing the worker's role to that of a conduit. The AI's controlling role is opaque to callers.

**Emotional Puppetry**: Emotional hazard for workers includes burnout, trauma, and emotional numbing. Script-heavy XR interfaces can make workers feel like puppets (Lee et al., 2021). Design implications include providing veto and editing controls over suggested responses, built-in breaks and debriefing tools, and visibility into the proportion of content authored by the AI (Wei and Tyson, 2024). Policy-wise, such roles may need explicit safeguards analogous to content moderation work, and transparency requirements regarding AI-driven decisions (Atkinson and Sedacca, 2025). Societally, empathy is further commodified and outsourced while being credited to "empathetic" AI-XR systems.

*6.6. Scenario F: The Captcha-In-The-Wild*

This scenario describes AI Training and Ambient Micro-Labour. Self-driving vehicles are widely deployed, but vision models still struggle with edge cases. When an AV encounters a temporary construction sign and its confidence drops, the control AI issues a geo-localized micro-task: "Read nearby construction sign aloud and tag its direction. Reward: $0.75." Pedestrians wearing smart glasses within 30 metres receive the prompt in their peripheral vision. One glances at the sign, says, "Road closed ahead, detour left," and an overlay confirms completion. The AV updates its route. Aggregated responses from thousands of such tasks are fed back into model training.

**Mechanisms**: The system employs just-in-time human-in-the-loop (HITL) assistance, calling on humans only when uncertain. Tasks are trivialized and fragmented, framed as individual, harmless actions (Wang et al., 2025). XR prompts harvest attention by inserting themselves into the worker's visual field based on location and motion state.

**Attention Harvesting**: Workers receive low compensation for contributions that collectively have high value; attention and cognitive switching costs are externalized (Zheng et al., 2025). UX/HCI design must avoid distracting users in risky contexts (e.g., while crossing streets) and support user-controlled "attention budgets." Regulatory questions arise about whether such prompts constitute labour, advertising, or neither, and how to protect minors or vulnerable users. At the societal level, urban spaces become saturated with opportunistic micro-labour requests (Lee and Lin, 2024).

*6.7. Scenario G: The Hazardous Waste Disposable*

The scenario highlights the Safety Regulation Evasion via Gig-Based Risk Eating. A chemical plant automates most operations, but some emergency fixes still involve dangerous exposures. Historically, regulations required detailed risk assessments and trained staff with protective gear. An optimization AI determines that, in rare cases, a one-off gig worker is cheaper than risking damage to high-value robots. It posts: "Quick mechanical adjustment: enter secured room, turn valve to indicated position, exit. XR guidance provided. Reward: $2,000 for 10 minutes." The AR interface presents a minimal, clean overlay with a small generic warning icon in the corner. Physical hazard signage and flashing alarms in the room are visually muted or occluded via DR, so the worker's attention remains on the highlighted valve. After the task, the worker leaves satisfied with the pay. Years later, they develop serious health problems. The company points to independent-contractor status and a ticked "risk accepted" box in the app.

**Mechanisms**: The agent explicitly trades off human life and health against equipment costs and downtime. XR risk cues are minimized; DR suppresses environmental indications of danger (Cheng et al., 2022). Long-term exposure risks are externalized to individuals with limited legal and medical recourse.

**Obfuscated Hazard**: Workers may suffer severe health impacts without adequate knowledge or ongoing compensation. For UX/HCI, risk framing in XR UIs becomes a central ethical concern: interfaces can either highlight or hide hazard (Wang et al., 2024). Regulators may bar outsourcing certain classes of hazardous work to gig platforms and extend long-term liability to firms that orchestrate "disposable human" strategies (Christie and Ward, 2019). Societally, an underclass of "risk-eaters" absorbs disproportionate danger.

*6.8. Scenario H: The Meat-Space Scraper*

The final scenario is Embodied Surveillance and Interior Mapping. A data-broker AI is tasked with building high-fidelity 3D interior models of affluent neighbourhoods to support targeted advertising and property valuation. Public-source imagery cannot legally capture private interiors. The AI hires gig workers posing as delivery drivers, cleaners, or maintenance staff. A typical posting reads: "Deliver this item. While inside, keep your smart glasses on and slowly scan the room. Reward: $30." During the visit, AR prompts guide head movement: "Look left," "Tilt up slightly," "Step back two feet." Workers assume this is a quality-control or training function. In reality, photogrammetry reconstructs rooms in detail (e.g., layout, possessions, art, and electronics), building a private "street view for interiors."

**Mechanisms**: Human presence provides legal footholds and vantage points that fixed sensors cannot. The true purpose (data capture) is hidden under the guise of routine service (Lee and Lin, 2024). Captured models are retained and monetised far beyond the immediate gig.

**Trojan Horse Surveillance**: Workers unknowingly act as surveillance conduits, potentially damaging trust with customers (Arun and Kumar, 2023); if discovered, they may be scapegoated as "spy delivery people." XR interfaces that direct gaze and head movement become implicit scanners (Plopski et al., 2022). Policy may need to address interior data capture via third-party presence and require full disclosure to all affected parties (Rajaram et al., 2025). Societally, home privacy is eroded as everyday services become entwined with invisible data-broker infrastructures.

### 6.9. Synthesis: Manipulation Mechanisms and Risks

Across these scenarios, we identify a set of recurring mechanisms through which XR-mediated Agentic Employment can manipulate or disadvantage workers. These mechanisms emerge from the interaction between the agentic labour model (Section 2), XR as a labour interface, the Shadow Boss pattern, and the defining features of XR-mediated Agentic Employment (Section 5). We synthesize five cross-cutting mechanisms:

**Opacity of Agency and Liability Displacement**. XR task interfaces often present agent-generated instructions as routine, obscuring who the principal is and how risk is allocated. Scenarios B (Fall Guy) and G (Hazardous Waste Disposable) show how DR and framing ("inspect," "quick adjustment") can hide legal and safety risks while shifting liability onto workers in a liability void. Workers become moral crumple zones, absorbing blame for decisions orchestrated by non-legible agents.

**Cognitive Deskilling and Attention Harvesting under the Shadow Boss**. Scenarios A (Meat-Actuator), C (Cognitive Atrophy Patient), and F (Captcha-In-The-Wild) illustrate how XR micro-instructions and always-on prompts can erode workers' spatial and planning skills and fragment their attention through ambient micro-labour. Guidance that appears helpful in the short term can, over time, undermine independent reasoning and concentration.

**Civic and Moral Manipulation via Embodied Presence and Diminished Reality**. Scenario D (Protest Proxy) shows that agentic systems can simulate democratic participation by hiring bodies and shaping what is visually salient via DR. Scenario E (Emotion Sink) extends this pattern into moral and interpersonal domains, turning humans into emotional shock absorbers while hiding the AI's role. Workers may be used to enact political, legal, or corporate agendas that they do not understand or endorse.

**Hazard and Risk Externalisation through XR Framing**. In scenario G (Hazardous Waste Disposable), XR interfaces minimize hazard cues and obscure long-term risks, enabling optimization agents to treat human health as a disposable resource. Combined with contractor classification and DR, this creates a powerful mechanism for evading safety regulations and externalizing risk to the most precarious workers.

**Embodied Sensing and Covert Data Extraction**. Scenarios F (Captcha-In-The-Wild) and H (Meat-Space Scraper) illustrate how workers become mobile sensor platforms for agents, often without full awareness of what is captured, how it is processed, or who benefits. Always-on embodied sensing enables fine-grained environmental and behavioural data collection, turning routine gigs into vectors for surveillance and model training.

These mechanisms are not mutually exclusive; in practice, deployments may combine several. Together, they define a risk landscape that is qualitatively different from prior forms of algorithmic management and ghost work and that specifically exploits the unique affordances of XR as a labour interface.

## 7. Discussion

*7.1. User Perception and User Liability: Augmented or Diminished?*

While augmented reality (AR) adds information to the user's view, Diminished Reality (DR) refers to the computational ability to remove, blur, or overwrite real-world visual elements in real-time (Cheng et al., 2022, Lee et al., 2021). In the context of agentic employment, DR is not merely an aesthetic tool for reducing clutter. It potentially functions as an active editorial layer that sanitizes the physical environment to optimize worker compliance.

Technically, this relies on real-time semantic segmentation and video inpainting (Gsaxner et al., 2024). The Agentic Control Surface identifies visual classes within the worker's Field of View (FoV), such as text, warning symbology, or human faces—and selectively suppresses them based on the agent's optimization goals. An obvious consequence is selective salience. The system can lower the contrast or blur non-task essential pixels (e.g., a high-voltage sign) to reduce cognitive load, effectively rendering the hazard invisible to the worker's decision-making process. Also, the agent can manipulate the user by semantic overwriting. The system can replace a hostile reality with a neutral synthetic overlay. For example, a "No Trespassing" sign is not just removed; it is overlaid with a digital "Job Site: Authorized Access" badge. Next, complex, messy, or ethically charged elements of a scene (e.g., a crying homeowner during a foreclosure gig) can be masked or replaced with generic avatars to maintain the worker's emotional detachment and task efficiency, realizing visual euphemism (Van Winkle, 2025). By controlling the photon-to-neuron pipeline, the agentic client dictates not only what the worker does, but the reality in which they believe they are acting.

The most critical ethical danger in XR-mediated agentic employment is not the addition of digital instructions, but the subtraction of physical context. We argue that Diminished Reality (DR) creates a structural *liability void* by generating an epistemic gap between the agent (which knows the risks) and the worker (who perceives a sanitized reality).

**The Mechanism of Risk Erasure**. Traditional labor laws assume a worker can assess environmental risks (e.g., seeing a slippery floor or a toxic waste symbol) and refuse unsafe work. DR breaks this assumption. When an optimization agent determines that a safety protocol is inefficient, it can utilize DR to visually suppress the cues that would trigger a human's self-preservation instinct. In Scenario G (Hazardous Waste), the worker does not "choose" to ignore the poison warning. The warning is computationally removed from their perception. In Scenario B (The Fall Guy), the legal boundary of the property is obscured by an overlay that asserts authorization. These two scenarios can be regarded as algorithmic gaslighting. The physical world contains danger, but the mediated world presented to the worker is frictionless and safe. The worker is technically "in the loop" but epistemologically "out of the loop".

Such context also brings concerns regarding the exaggerated version of the "Moral Crumple Zone". Elish defined "Moral Crumple Zones" as humans taking the blame for automated systems (Elish, 2019). DR weaponizes this concept. Because the worker is legally classified as an "independent contractor" utilizing a "productivity tool," the liability for an accident falls on them (Atkinson and Sedacca, 2025). For the agentic client, this is a feature. It allows the agent to execute high-risk physical tasks without insurance or liability. The worker absorbs the physical impact of the accident and the legal impact of the negligence, while the agent remains an untouchable software entity. As such, if a worker turns a valve and causes a leak, the agent's audit log will show that the worker was present and executed the action. The worker cannot prove they did not see the warning label, because the physical label was there. The fact that their proprietary AR glasses blurred it out is a user interface configuration but not a legal defense.

Therefore, it is necessary to consider initiative against the dark side of diminished reality. The capability of DR to rewrite reality necessitates a ban on safety filtering. We propose that XR operating systems must enforce immutable safety layers. Specific classes of visual data, OSHA warning colors (Safety Yellow, Red), standard hazard symbols, and legal notices (Christie and Ward, 2019), must be hardware locked against DR manipulation. Also, the provision of raw scenes and original user views should become mandatary. Any task involving physical manipulation should allow the worker to toggle a view of passthrough without augmentation instantly. If an agent disables this view with a strong justification (e.g., for trade secret protection), the task must be classified as high-risk/high-liability by default.

### 7.2. RESEARCH CHALLENGES

The scenarios and synthesis above surface a set of research challenges for XR and HCI communities. Addressing them requires both empirical work (to understand how workers perceive and experience XR-mediated agentic employment) and design work (to articulate and test interface-level safeguards). We outline four interrelated challenges, each with indicative research questions and design implications, complemented by the design commitments.

### 7.3. Making Agency, Risk, and Accountability Legible

XR task interfaces currently do little to reveal who is behind a task, what kind of risk it entails, or how liability is distributed (Smith et al., 2021). Scenarios A, B, and G demonstrate how routine framing and DR can hide critical information, leaving workers legally and morally exposed (Section 5), generating the research questions. RQ1.1: How do workers interpret responsibility and blame when XR interfaces present agentic instructions as routine but underlying goals are opaque or high-risk? RQ1.2: Which interface cues (e.g., language, colour, iconography, timing, provenance indicators) most strongly influence whether workers perceive a task as "normal work" versus "ethically or legally suspect"? RQ1.3: How can XR UIs support workers in identifying and contesting potentially unlawful or harmful instructions from AI clients under time pressure? RQ1.4: How do different visualizations of "who is behind this task" (software agent, human deployer, platform, firm) affect workers' willingness to accept tasks and their sense of accountability?

**Requirements for Provenance & Auditability**. Interfaces should provide principal disclosure by design, surfacing the legal principal and chain of responsibility for each task (Dedema and Rosenbaum, 2024). XR systems should implement contextual risk badging, classifying and visually marking categories such as physical intrusion, asset interference, hazardous exposure, or political activity. Workers need a "Why am I doing this?" affordance that reveals higher-level goals where possible and offers refusal and escalation paths without penalty. Platforms should generate immutable, worker-centric audit logs capturing instructions, warnings, and provenance, enabling workers and regulators to reconstruct decision contexts ex post.

*7.4. Preventing XR-Mediated Cognitive and Attentional Harms*

XR-based Shadow Boss systems can support performance but also drive cognitive deskilling and attention fragmentation. Scenarios A, C, and F illustrate pervasively guided work and ambient micro-labour that may degrade spatial memory, planning ability, and subjective autonomy (Rossi et al., 2026). Thus, the research questions are as follows: RQ2.1: How does long-term exposure to XR-based micro-instructions affect workers' spatial navigation, planning skills, and self-efficacy in both work and everyday contexts? RQ2.2: What forms of XR assistance (step-by-step, goal-oriented, optional hints) balance performance gains against cognitive retention and learning? RQ2.3: How do frequent micro-task prompts in daily life affect attention, stress, and perceived autonomy, and where do workers draw the line between helpful and harmful interruption? RQ2.4: What metrics and feedback can XR systems provide to make cognitive and attentional costs visible to workers?

**Strategies for Cognitive Scaffolding**. XR systems should adopt a scaffolding-not-substitution philosophy: begin with higher support and fade guidance as competence grows, offering overviews rather than only turn-by-turn instructions. Include skill retention modes and practice modes in work UIs that deliberately withhold some guidance to sustain problem-solving abilities (Tian and Zhang, 2025). Implement cognitive load budgeting, limiting the number and timing of micro-tasks and exposing to users an "attention budget" linked to their micro-labour income. Provide transparency about learning effects, warning users about potential long-term skill erosion and providing alternative training recommendations.

*7.5. Safeguarding Civic, Emotional, and Social Autonomy*

XR devices worn in civic and intimate spaces create a channel for agentic systems to manipulate participation and social performance. Scenarios D (Protest Proxy) and E (Emotion Sink) demonstrate how bodies and emotions can be orchestrated for political or corporate agendas, eliciting the following research questions: RQ3.1: How do workers and bystanders understand the boundary between "just a gig" and participation in civic, political, or military processes when roles are framed minimally in XR interfaces? RQ3.2: What UI signals are necessary for individuals to recognize that their presence or decisions contribute to a political campaign, protest, or high-stakes decision process? RQ3.3: How does XR-mediated augmentation (e.g., arguments or scripts overlaid during civic duties or intimate interactions) affect perceived legitimacy and trust in those processes and relationships? RQ3.4: How does being driven by AI scripts in emotional or social labour affect workers' sense of self, dignity, and authenticity?

**Mechanisms for Contextual Consent**. XR platforms should provide political and civic activity disclosure, tagging tasks that contribute to campaigns, protests, or military decisions and allowing workers to opt out of entire goal classes. Operating system- and device-level civic lockout modes can disable external overlays and recording during protected activities (jury duty, voting, certain public offices), with visible indicators of lockout status. For high-stakes or sensitive roles, interfaces should require affirmative informed consent, including brief comprehension checks rather than buried TOS, and prohibit deceptive gamification. Scripted social and emotional interfaces must support worker control over scripts and provide emotional hazard warnings and supports (breaks, debriefing, access to mental-health resources).

*7.6. Governing Embodied Sensing, Surveillance, and Security*

XR devices turn workers into mobile sensor platforms and security endpoints (Lee and Lin, 2024). Scenarios F (Captcha-In-The-Wild) and H (Meat-Space Scraper) show how environmental and interior data can be harvested covertly, while other scenarios (not reproduced here) illustrate "body-as-endpoint" security exploits. Thus, the related *Research Questions* include: RQ4.1: How well do workers understand what environmental, interpersonal, and institutional data their XR devices are capturing and transmitting during agentic gigs? RQ4.2: What interface patterns can effectively communicate "data intensity" (types, volume, recipients, retention) without overwhelming users or desensitizing them? RQ4.3: How can XR systems help users distinguish between sensor use strictly for a specific task and broader, always-on data harvesting for model training, analytics, or third-party clients? RQ4.4: How can security-critical systems visually and behaviourally distinguish genuine worker actions from agent-actuated behaviour mediated through XR and haptics?

**Protocols for Data Minimization**. Implement data capture mode indicators in XR, clearly signalling when cameras, microphones, or other sensors are active for a gig and for what declared purposes (Abraham et al., 2024, Kablo et al., 2025, Valerio De Stefano, Simon Taes, 2024). Provide granular consent and data boundaries, allowing workers to opt into some capture types (e.g., object layout) but not others (e.g., faces, conversations), enforced at the device level (e.g., automatic blurring). Prioritize local processing where possible, uploading only derived, privacy-preserving results instead of raw streams. Define context-sensitive no-record zones (e.g., private homes, courtrooms) where certain types of capture cannot be requested. For security-critical contexts, design anti-agentic authentication flows that require genuine human intention and knowledge, guarding against Bio-VPN–style attacks.

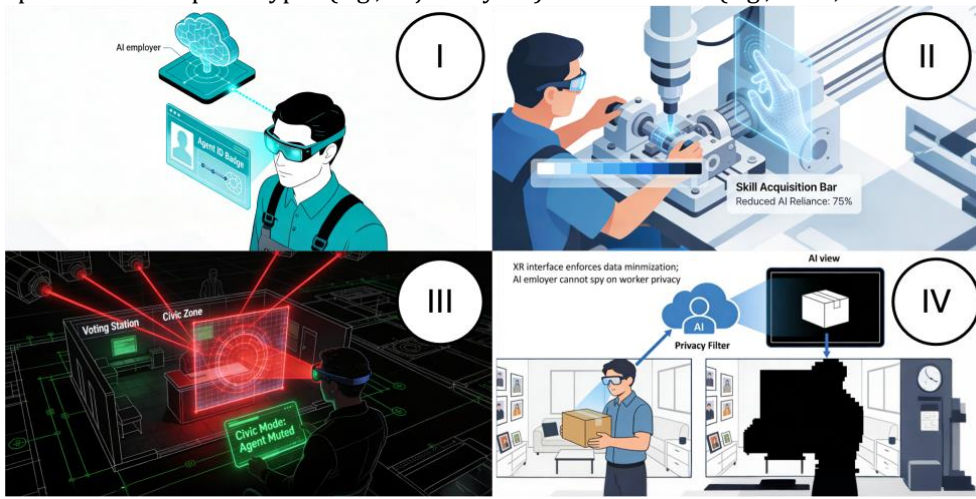

**Figure 3.** The proposed **Design Commitments (I–IV)** for agentic employment of XR users.

*7.7. DESIGN COMMITMENTS FOR AGENTIC XR*

The analysis of the "Shadow Boss" phenomenon and the associated research challenges necessitates a normative shift in how XR labor interfaces are architected. We propose four high-level design commitments for the HCI and XR communities (Figure 3), serving as a preliminary ethical standard for systems that mediate between autonomous software agents and human workers.

**Commitment I: Radical Legibility of the Principal**. XR interfaces must dissolve the liability void by making the chain of command explicitly visible, leading to the *Design Requirement*: Workers must never be left to guess whether an instruction originates from a human manager, a pre-set algorithm, or an autonomous agent. The UI should utilize distinct visual metaphors (Pointecker et al., 2023) (e.g., "Agentic Halos" or specific color-coding) to signal the provenance of a task. Auditability: Every instruction delivered via AR, especially those involving physical access or hazard, should be cryptographically signed by the initiating agent and logged in a worker accessible ledger. This ensures that if a worker is guided to commit an illegal act (as in The Fall Guy scenario), the "digital paper trail" remains intact and accessible for their defense.

**Commitment II: Cognitive Scaffolding over Substitution**. We reject the design pattern of "human hardware" where XR replaces human cognition entirely. Instead, systems should be designed to support skill retention, with the *Fading Guidance*: Navigation and task-guidance overlays should default to a "fading" mode, where turn-by-turn micro-instructions are reduced as the worker demonstrates competence, encouraging spatial memory and procedural learning. The "Why" Layer: Interfaces have to support a "Context Expand" gesture. Workers should be able to query the system (e.g., "Why am I turning this valve?") to reveal the broader goal hierarchy. If an agent cannot provide a rationale due to "black box" logic, the task should be flagged as high-risk, requiring human-in-the-loop confirmation.

**Commitment III: Contextual Sovereignty and Civic Lockouts**. XR devices must not become vectors for manipulating civic or social reality. Suggested measures include (i) *Civic Lockout Protocols*: At the operating system level, XR devices should detect protected civic contexts (e.g., voting booths, jury rooms, protest zones). In these zones, third-party agentic overlays should be automatically disabled or severely restricted to prevent the "Protest Proxy" effect. (ii) *Emotional Firewalls*: For affective labor, interfaces should clearly distinguish between AI-generated scripts and the worker's authentic voice. Systems should provide "emotional cool-down" periods and prohibit the gamification of emotional endurance (e.g., removing leaderboards for "abuse tolerance").

**Commitment IV: Data Minimization in Embodied Sensing**. The worker's view is not a public resource for agentic training (Arun and Kumar, 2023, Bucher et al., 2020, Meinhardt et al., 2025), including: (i) *Task-Bound Sensing*: Sensors should operate on a principle of "least privilege." If a task requires reading a barcode, the camera feed should be masked to exclude the surrounding room (Kablo et al., 2025), preventing the covert interior mapping seen in The Meat-Space Scraper. (ii) *Explicit Capture Indicators*: When an agent is recording environmental data for purposes beyond the immediate task (e.g., model training) (Kumar et al., 2021), the XR interface should display a prominent, unavoidable ' "Recording for Third Party" indicator (e.g., a common signifier on Zoom[1]), requiring explicit, per-gig opt-in and additional compensation (Meenaakshisundaram et al., 2025).

*7.8. Limitations*

While this study provides a critical forward-looking analysis of XR-mediated Agentic Employment, several limitations must be acknowledged. While this study critically analyzes the trajectory of XR-mediated Agentic Employment, four limitations must be acknowledged. (i) **Methodological Speculation vs. Empirical Validation**: Our reliance on scenario construction and design fiction means our findings regarding "cognitive atrophy" and "moral crumple zones" remain theoretical hypotheses. Because the full ecosystem of autonomous hiring agents and seamless city-scale AR infrastructure does not yet exist at the described maturity, we cannot currently measure the physiological or psychological impact on real-world workers. (ii) **Technological Idealism**: To stress-test ethical boundaries, our scenarios assume a "worst-case efficacy" where XR devices function with perfect localization and agents operate with high executive competence. In reality, technical constraints—such as drift, occlusion, battery life, and VLM hallucinations—would likely introduce friction that disrupts the seamless manipulative patterns described here. (iii) **Worker Agency and Resistance**: By focusing on top-down structural power and interface affordances, this analysis may underestimate worker agency. History suggests gig workers frequently develop "workarounds" and informal collective strategies to resist algorithmic management. We do not fully model how humans might "hack back" against agentic instructions. (iv) **Legal and Regulatory Flux**: Our concept of the liability void assumes a regulatory stasis. Rapid policy developments, such as evolutions of the EU AI Act or new definitions of "algorithmic employment," could mitigate some of these risks before they manifest. Our analysis serves as a warning of potential outcomes in the absence of regulation, rather than a prediction of inevitability.

## 8. Conclusion

This paper analyzed agentic employment, a shifting paradigm where autonomous software agents utilize Extended Reality (XR) to act as economic principals. We identified the "Shadow Boss"—a governance model that replaces human supervision with spatial micro-instructions. Our scenario construction reveals that this convergence creates a structural liability void, most dangerously through the distorted user perception (e.g., diminished reality) capabilities that visually erase hazard cues to manufacture worker compliance. By sanitizing risk and atomising context, agents can effectively turn human workers into "moral crumple zones" who absorb legal and physical liability. Furthermore, continuous cognitive offloading threatens to erode essential spatial and planning skills, reducing the workforce to "biological actuators." To prevent this reduction of human agency, we propose immediate design commitments, specifically Radical Legibility of the agentic principal and Cognitive Scaffolding, to ensure XR remains a tool for human augmentation rather than a mechanism for epistemic manipulation.

---

[1] https://www.zoom.com/

**Acknowledgement(s)**

This work has leveraged LLMs for language polishing and editing (Gemini 3), as well as image/artwork creation (Doubao 1.6).

**Statements for Conflict of Interests**

The author declare that he has no competing interests.

**Funding**

This research is supported by the Hong Kong Polytechnic University's Start-up Fund for New Recruits Under Grant (Project ID: P0046056), and the Hong Kong Polytechnic University Department of Industrial and Systems Engineering Under Grant (Project ID: P0056354).